\documentclass[12pt,reqno,a4paper,twoside]{article}
\usepackage{amsmath,amsthm,amstext,amscd,amssymb,euscript}
\usepackage{epsf}
\renewcommand{\phi}{\varphi}

\renewcommand{\emptyset}{\varnothing}

\newcommand{\Zd}{\mathbb Z^d}

\def\1{ {\mathit{1} \!\!\>\!\! I} }

\newcommand{\ind}{1\hspace{-2.6mm}{1}}

\makeatletter
\@addtoreset{equation}{section}
\makeatother

\newtheorem{ittheorem}{Theorem}
\newtheorem{itlemma}{Lemma}
\newtheorem{itproposition}{Proposition}
\newtheorem{itdefinition}{Definition}
\newtheorem{itremark}{Remark}

\newenvironment{theorem}{\addtocounter{equation}{1}
\begin{ittheorem}}{\end{ittheorem}}

\newenvironment{lemma}{\addtocounter{equation}{1}
\begin{itlemma}}{\end{itlemma}}

\newenvironment{proposition}{\addtocounter{equation}{1}
\begin{itproposition}}{\end{itproposition}}

\newenvironment{definition}{\addtocounter{equation}{1}
\begin{itdefinition}}{\end{itdefinition}}

\newenvironment{remark}{\addtocounter{equation}{1}
\begin{itremark}}{\end{itremark}}

\newenvironment{corollary}{\addtocounter{equation}{1}
\begin{itcorollary}}{\end{itcorollary}}

\newcommand{\beq}{\begin{eqnarray}}
\newcommand{\eeq}{\end{eqnarray}}

\newcommand{\be}{\begin{equation}}
\newcommand{\ee}{\end{equation}}

\newcommand{\bl}{\begin{lemma}}
\newcommand{\el}{\end{lemma}}

\newcommand{\br}{\begin{remark}}
\newcommand{\er}{\end{remark}}

\newcommand{\bt}{\begin{theorem}}
\newcommand{\et}{\end{theorem}}

\newcommand{\bd}{\begin{definition}}
\newcommand{\ed}{\end{definition}}

\newcommand{\bp}{\begin{proposition}}
\newcommand{\ep}{\end{proposition}}

\newcommand{\bc}{\begin{corollary}}
\newcommand{\ec}{\end{corollary}}

\newcommand{\bpr}{\begin{proof}}
\newcommand{\epr}{\end{proof}}

\newcommand{\bi}{\begin{itemize}}
\newcommand{\ei}{\end{itemize}}

\newcommand{\ben}{\begin{enumerate}}
\newcommand{\een}{\end{enumerate}}

\newcommand{\Z}{\mathbb Z}
\newcommand{\R}{\mathbb R}
\newcommand{\N}{\mathbb N}

\newcommand{\E}{\mathbb E}

\newcommand{\pee}{\ensuremath{\mathbb{P}}}
\newcommand{\re}{\ensuremath{\mathcal{R}}}

\newcommand{\haa}{\ensuremath{{\bf \mathcal H}}}

\newcommand{\s}{\ensuremath{\mathcal{S}}}

\newcommand{\fe}{\ensuremath{\mathcal{F}}}

\newcommand{\la}{\ensuremath{\Lambda}}
\newcommand{\si}{\ensuremath{\sigma}}

\def\now{
\ifnum\time<60
          12:\ifnum\time<10 0\fi\number\time am
          \else
            \ifnum\time>719\chardef\a=`p\else\chardef\a=`a\fi
          \hour=\time
          \minute=\time
          \divide\hour by 60 
          \ifnum\hour>12\advance\hour by -12\advance\minute by-720 \fi
          \number\hour:%
          \multiply\hour by 60 
          \advance\minute by -\hour
          \ifnum\minute<10 0\fi\number\minute\a m\fi}
\newcount\hour
\newcount\minute
\numberwithin{equation}{section}         

\theoremstyle{remark}

\newcommand{\caP}{{\mathcal P}}

\begin{document}
\title{{\bf Organized versus self-organized criticality in the abelian sandpile model}}
\author{A. Fey-den Boer\footnote{Faculteit exacte wetenschappen, Vrije Universiteit Amsterdam,
De Boelelaan 1105, 1081 HV Amsterdam and EURANDOM,
Technische Universiteit Eindhoven, Postbus 513,
5600 MB Eindhoven, The Netherlands, afey@eurandom.nl}
\\
F.\ Redig\footnote{Faculteit Wiskunde en Informatica, Technische Universiteit Eindhoven, Postbus 513,
5600 MB Eindhoven, The Netherlands, f.h.j.redig@tue.nl}\\
\\
}
\maketitle
\footnotesize
\begin{quote}
{\bf Abstract:} We define stabilizability of an infinite volume height configuration and of
a probability measure on height configurations. We show that for high enough densities,
a probability measure cannot be stabilized. We also show that in some sense the thermodynamic limit
of the uniform measures on the recurrent configurations of the abelian sandpile model (ASM)
is a maximal element of the set of stabilizable measures. In that sense the self-organized
critical behavior of the ASM can be understood in terms of an ordinary transition between stabilizable
and non-stabilizable.
\end{quote}

\section{Introduction}

Self-organized criticality (SOC) is a concept introduced in \cite{BTW} to model power-law behavior of
avalanche sizes in various natural phenomena such as sand and rice piles, forest fires, etc.
The conceptual point of view in \cite{BTW} is that this kind of criticality is not tuned by parameters such as
temperature or magnetic field, as is the case in critical systems of equilibrium statistical
mechanics. This point of view has been
questioned by several people, see e.g.\ \cite{dick,ala}, where it is argued that
the choice of the models exhibiting SOC involves an {\em implicit}
tuning of parameters, and hence SOC is an (interesting) example of ordinary criticality. In the case of the abelian sandpile model, e.g. one can say that the choice of the toppling
matrix (which governs the dynamics) having mass equal to zero is a fine tuning. Indeed in the massive or  dissipative case
(where in the bulk grains are lost upon toppling) the avalanche sizes exhibit exponential decay, so in
that case there is no criticality.

Similarly, in \cite{quant}, the authors investigate the relation between the critical density
of some parametric model of random walkers with that of the abelian sandpile model, and
prove in $d=1$ that ASM density corresponds exactly to the transition point in the random
walkers model. They further conjecture that this is also true in $d\geq 2$.
In this paper we want to continue the relation between an ordinary critical phenomenon and
the SOC-state of the abelian sandpile model.  This is done through the notion of
``stabilizability". A height configuration is called stabilizable if upon stabilizing  it in larger and
larger volumes, the number of topplings at a fixed site does not diverge. This implies that we can
``redistribute" the mass in {\em infinite} volume such that after the redistribution, all sites
have a height between $1$ and $2d$.
Similarly a probability measure $\nu$ on height configurations is called stabilizable
if it concentrates on the set of stabilizable configurations.
The conjecture in \cite{quant}, inspired by \cite{dick} is that there exists
$\rho_c>0$ such that (modulo some restrictions on the measure $\nu$) if the $\nu$ expected height $\rho<\rho_c$, then
$\nu$ is stabilizable, if $\rho>2d$ it is not stabilizable, and
for any $\rho\in (\rho_c,2d)$ there exist measures $\nu$ with
expected height $\rho$ which are not stabilizable. Moreover, $\rho_c$ is exactly the expected height in the stationary state
of the abelian sandpile model, in the thermodynamic limit.

The aim of this paper is to prove the last two items of this conjecture, and to give some
more insight in the regime $\rho\in (\rho_c, 2d)$.
Our paper is organized as follows: in section 2 we define the notion of stabilizability. In section
3 we precisely state the main conjecture of \cite{quant} and prove item 3 of it. In section
4 we prove item 2 of the conjecture and disprove by a counterexample item 1. We further show that in some sense the infinite volume limit
of the stationary measure of the abelian sandpile model is ``maximal stabilizable".
Finally in section 5 we introduce the concept of meta-stabilizability, and give a class of examples
of probability measures on height configurations having this property.

\section{Basic definitions}
A height configuration is defined as a map
$\eta: \Zd\to \{1,2,\ldots\}$, where by convention the minimal height is chosen to be 1.
The set of all height configurations is denoted by $\haa$. The set of
probability measures on the Borel sigma-field of $\haa$ is denoted $\caP (\haa)$.
A configuration is called {\em stable} if for all $x\in\Z^d$, $\eta (x)\leq 2d$. The set
of stable configurations is denoted by $\Omega$. Similarly, for $V\subset \Zd$,
$\Omega_V$ denotes the
set of stable configurations $\eta: V\to \{ 1,\ldots 2d\}$.

For $V\subset \Z^d$ the abelian sandpile toppling matrix is defined as
\be\label{delta}
(\Delta_V)_{x,y} = 2d\delta_{x,y} - \ind_{x,y\in V, \ |x-y|=1}
\ee
Its inverse is denoted by
\be\label{green}
G_V (x,y) = (\Delta_V^{-1})_{x,y}
\ee
The probabilistic interpretation of $G_V$ is:
\be
G_V (x,y) = \frac{1}{2d}\E^V_x \left(\mbox{number\ of\ visits\ at\ site}\ y\right)
\ee
where $\E^V_x$ denotes expectation in the simple random walk started at $x$, and killed
upon exiting $V$.
\bd\label{stab}
A height configuration $\eta\in\haa$ is called stabilizable if for
any sequence of volumes $V_n\uparrow\Zd$, there exist $m_{V_n}\in \N^{V_n}$ such that
\be
\eta_{V_n}-\Delta_{V_n} m_{V_n} =\xi_{V_n} \in \Omega_{V_n}
\ee
and for any $x\in\Z^d$ $m_{V_n} (x) \to m(x)$ as $n\to\infty$.
\ed
The set of all stabilizable configurations is denoted by $\s$. It follows immediately from the definition that
for $\eta\in \s$, and $m= \lim_{n\to\infty} m_{V_n}$,
\be\label{weakstab}
\eta-\Delta m = \xi\in\Omega
\ee
where $\Delta_{x,y} = 2d\delta_{xy} - \ind_{|x-y|=1}$ is the infinite volume toppling matrix.

\bd\label{stabdef}
A probability measure $\nu$ on height configurations is called stabilizable if $\nu (\s) =1.$
\ed
It is clear that the set of stabilizable configurations
is a translation invariant subset of $\haa$. Therefore any stationary and ergodic probability
measure $\mu$ on $\haa$ satisfies $\mu (\s) \in \{ 0,1\}$.

Remark that in finite volume $V\subset\Zd$, the abelian sandpile model is ``well-defined". This
means that for any height configuration $\eta\in\N^V$, the equation
\be\label{kwaak}
\eta_V - \Delta_V m_V =\xi_V
\ee
with unknowns
the couple $(m_V, \xi_V)$ has at least one solution, namely
for $x\in V$, $m_V(x)$ equals the number of topplings at $x$ needed to stabilize
$\eta$ in $V$ (see e.g. \cite{MRZ}).
Notice that the couple $(m_V, \xi_V)$ is not unique, but if $m_V$ is such
that $\xi_V$ is stable and (\ref{kwaak}) holds, then $m_V (x)$ is  {\em at least}
the number of
topplings at $x$ needed to stabilize $\eta_V$. So the vector collecting
the number of topplings (needed to stabilize $\eta_V$)
is the {\em minimal} solution $m_V$ of the equation (\ref{kwaak}). We will always
choose this solution in the sequel.

On $\haa$ we have the natural pointwise ordering $\eta\leq\xi$ if $\forall x\in\Z^d$, $\eta (x)\leq \xi (x)$.
A function
$f:\haa\to\R$ is called monotone if $\eta\leq\xi$ implies $f(\eta) \leq f(\xi)$.
Two probability measures $\mu,\nu$ on $\haa$ are ordered $\mu\leq \nu$ if for any
monotone function,
$\int f d\mu\leq \int f d\nu$. This is equivalent to the existence of a coupling
$\pee$ of $\mu$ and $\nu$ such that $\pee (\{ (\eta,\xi): \eta\leq \xi\}) =1$.

By abelianness, $m_V (x)$ is non-decreasing in $V$. Therefore a configuration
$\eta$ is not stabilizable if and only if there exists $x\in\Z^d$ such that
$m_V (x)\uparrow\infty$. By abelianness, the $m_V$ are also monotone functions
(in the sense just mentioned) of the configuration
$\eta$.

Therefore we have the following immediate properties of the set of stabilizable configurations
\bp\label{els}

\begin{itemize}
\item[a)]
$\s$ is a translation invariant measurable set.
\item[b)]
If $\eta\in\s$ and $\xi\leq\eta$, then $\xi\in\s$.
\item[c)] If $\mu$ is a stabilizable probability measure, and
$\nu\leq\mu$, then $\nu$ is a stabilizable probability measure.
\end{itemize}
\ep
We then define the following ``critical densities":
\bl\label{monlem}
Define
\beq
\rho_c^+ &=& \inf \{ \rho\geq 1: \ \exists\nu\in\caP (\haa)
\ \mbox{with} \ \nu (\eta (0))=\rho,\ \mbox{and}\ \nu\ \mbox{is not
stabilizable}\}
\nonumber\\
\rho_c^-  &=& \sup\{\rho\geq 1: \ \forall\nu\in\caP (\haa)
\ \mbox{with} \ \nu (\eta (0))=\rho,\ \nu\ \mbox{is
stabilizable}\}
\eeq
Then $\rho^+_c =\rho^-_c$.
\el
\bpr
It suffices to see that the set
\be\label{set}
S=\{\rho\geq 1: \mbox{ such that}\ \forall\nu\in\caP (\haa)
\ \mbox{with} \ \nu (\eta (0))=\rho,\ \nu\ \mbox{is
stabilizable}\}
\ee
is an interval. Suppose that $\rho\in S$ and $\rho'<\rho$.
Consider a measure $\nu'\in\caP(\haa)$ with $\nu' (\eta (0))=\rho'$. Then there exists
a measure $\nu\in\caP (\haa)$ such that $\nu(\eta (0))=\rho$ and
$\nu\geq\nu'$. Since $\nu$ is stabilizable, by the monotonicity property \ref{els}
item 3, $\nu'$ is stabilizable.
\epr

We now introduce the ``critical state" of the sandpile model, and its thermodynamic
limit. Define a configuration allowed in a volume $V$ if
for any subset $W\subset V$,
the inequality
\be
\eta (x) \leq |\{y\in W,|y-x|=1\}|
\ee
is violated for at least one $x\in W$. The set of allowed configurations in volume
$V$ is denoted by $\re_V$. It is well known that the set of recurrent configurations
of  the abelian
sandpile model in finite volume $V$ coincides with the set of allowed configurations $\re_V$, and
the stationary measure is the uniform measure on $\re_V$.
(see e.g.\ \cite{MRZ} or the basic reference \cite{dhar}). We denote
this measure by $\mu_V$. Recently, it has been proved
in \cite{MRS, AJ} that the weak limit $\mu =\lim_{V\uparrow\Z^d} \mu_V$ exists and
defines a measure on infinite volume height configurations. Moreover, its support
$\re$ is the set of those configurations $\eta$ such that all restrictions to finite volumes $V$ have the
property $\eta_V\in\re_V$. We will call this measure
$\mu$ the uniform measure on recurrent configurations (UMRC). We will always
use the symbol $\mu$ for the UMRC and denote its mean height by
$\rho_c= \mu (\eta (0))$. In \cite{AJ} it is proved that $\mu$ is translation invariant, while
in \cite{JR} it is proved that $\mu$ is tail-trivial so in particular ergodic under spatial
translations.

\section{Main conjecture and results}
In the rest of the paper we will prove point 2 and 3 of the following conjecture appearing
in \cite{quant}, cf.\ also \cite{dick}, and
we will also give some additional new results and examples.
\\
{\bf Conjecture:}
\\
Let $\nu$ be a stationary and ergodic probability measure on $\N^{\Zd}$.
Put $\rho = \nu (\eta (0))$.
\ben
\item
For $\rho <\rho_c$, $\nu$ is stabilizable
\item
For $\rho_c< \rho \leq 2d$ there exist $\nu$ which are
not stabilizable
\item
For $\rho> 2d$, $\nu$ is not stabilizable
\een

The following theorem settles point 3 of the conjecture.
\bt\label{mainprop}
Suppose that $\eta$ has a distribution $\nu$  such that
$\nu (\eta(0)) =\rho >2d$. Then $\eta$ is almost surely not stabilizable.
\et
\bpr
Let $G_V$ be the Green function  introduced in (\ref{delta}), (\ref{green}).
Simple random walk killed upon exiting $V$ will be denoted by $\{ X_n, n\in\N \}$, and corresponding expectation
by $\E^V_0$. Finally, let $\tau_V$ denote the lifetime of this walk. The infinite volume random walk
expectation is denoted by $\E_0$. Of course, $\E_0$ and $\E_0^{V}$-expectation of events
before $\tau_V$ coincide.

Suppose that $\eta$ drawn from $\nu$ is stabilizable.
Then we have
\beq\label{schatt}
m_V (0)& = & \sum_{x\in V} G_V (0,x) (\eta (x) -\xi(x))
\nonumber\\
&= &
(2d)^{-1}\E_0 \left( \sum_{n=0}^{\tau_V} (\eta (X_n) -\xi (X_n))\right)
\eeq
and $m_V(0)\uparrow m(0)<\infty$ as $V\uparrow\Zd$.
Since the random field $\eta$ is stationary and ergodic, we have
\[
\lim_{V\uparrow\Z^d} \frac1{\tau_V}\sum_{n=1}^{\tau_V} \eta (X_n) = \rho
\]
$\pee_0\times\nu$ almost surely, where $\pee_0$ denotes the path-space measure
of the simple random walk starting at $0$. For $\xi$ we cannot conclude such a strong statement but
we have, by stability
\[
\limsup_{V\uparrow\Z^d} \frac1{\tau_V}\sum_{n=1}^{\tau_V} \xi (X_n)\leq 2d
\]
Therefore since $\rho > 2d+\delta$ for some $\delta>0$,
\be
\liminf_{V\uparrow\Z^d} \frac1{\tau_V}\sum_{n=1}^{\tau_V}( \eta (X_n)- \xi(X_n)) >\delta
\ee
This implies, using Fatou's lemma, and the fact $\tau_V\to\infty$ as $V\uparrow\Zd$,
that for any $A>0$
\beq
&&2d\liminf_{V\uparrow\Zd} m_V (0)= \liminf_{V\uparrow\Z^d}\E_0 \left( \sum_{n=0}^{\tau_V} (\eta (X_n) -\xi (X_n)\right)
\nonumber\\
&\geq &
\liminf_{V\uparrow\Z^d}\E_0\left( A \ind_{(\tau_V >A)} \frac1{\tau_V}\left( \sum_{n=0}^{\tau_V} (\eta (X_n) -\xi (X_n))\right)\right)
\nonumber\\
&\geq &
\E_0\left( \liminf_{V\uparrow\Z^d}A \ind_{(\tau_V >A)}
\liminf_{V\uparrow\Z^d}\frac1{\tau_V}\left( \sum_{n=0}^{\tau_V} (\eta (X_n) -\xi (X_n))\right)\right)
\nonumber\\
&\geq &
A\delta
\eeq
Since $A>0$ is arbitrary, we arrive at a contradiction.

\epr
\section{Adding to the stationary measure}
In this section we settle point 2 of the conjecture.

The UMRC $\mu$ is obtained as a limit of finite volume stationary
measure $\mu_V$. These $\mu_V$ are in turn obtained by running the finite
volume addition and relaxation process for a long time.
Therefore, one can believe that $\mu$ is ``on the edge" of stabilizability.
More precisely if one could still ``add mass" to $\mu$, then $\mu$ would
not be stationary.

However, it is not true that $\mu$ is a maximal stabilizable measure
in the sense of
the FKG ordering of measures. Indeed, one can create the following
translation invariant $\nu$: pick a configuration according to $\mu$ and flip
all the height ones to height four. This measure is strictly dominating
$\mu$ in FKG sense, but it concentrates on stable configurations.
In the last section of this paper we will show that such ``artificially stable"
measures are in some sense ``metastable".

The idea of formalizing the maximality of $\mu$ is that ``one cannot
add mass to $\mu$".
For $\mu$ a probability measure on $\haa$, and $\nu$ a probability measure on
$\N^{\Z^d}$, we denote by
$\mu\oplus\nu$ the distribution of $\eta+\xi$ where $\eta$ is distributed according
to $\mu$ and $\xi$ is {\em independent} of $\eta$ and distributed according to $\nu$.

\bd
A probability measure $\mu$ on $\haa$ is called maximal stabilizable if for any
ergodic translation invariant $\nu$ with $\nu (\eta (0)) >0$,
$\mu \oplus\nu$ is not stabilizable.
\ed
In order to state our main result of this section, we need some
more conditions on the UMRC $\mu$. For a configuration drawn from $\mu$, define
the addition operator $a_{x,V}$ by
$a_{x,V} (\eta) = (a_{x,V}\eta_V)\eta_{V^c}$. In words this means that application
of $a_{x,V}$ to a configuration in $V$, te configuration changes
as if we added at $x$ and stabilized in $V$, while outside $V$ the configuration remains unaltered.

We say that the infinite volume addition operator $a_x = \lim_{V\uparrow\Zd} (a_{x,V})$ is
well-defined w.r.t.\ the UMRC if for $\mu$ almost every $\eta$, the limit
$\lim_{V\uparrow\Zd}a_{x,V} (\eta)$ exists (in the product topology). We now
can state our conditions
\bd\label{canon} The UMRC is called canonical if
\ben
\item
The infinite volume addition operators $a_x$ are well-defined w.r.t.\ the UMRC for any
$x\in\Zd$.
\item
The UMRC is stationary w.r.t.\ the action of $a_x$, i.e., if $\eta$ is distributed according to
the UMRC, then so is $a_x\eta$.
\een
\ed
In \cite{JR} we prove that these conditions are satisfied on $\Zd, \ d\geq 5$. The restriction
$d\geq 5$ is however of a technical nature, and we strongly believe that these
conditions are satisfied as soon as $\mu$ exists.
If the
UMRC $\mu$ is canonical, then one can easily see that finite products of
addition operators are well-defined $\mu$ a.s.\ and leave $\mu$ invariant.
See \cite{JR} for a complete proof.

Our main result in this section is the following.
\bt
If the  UMRC  is canonical, then it is maximal stabilizable.
\et
\bpr
We have to prove that $\mu \oplus\nu$ is not stabilizable for
any $\nu$ stationary and ergodic such that $\nu (\eta (0)) >0$. A configuration
drawn from $\mu\oplus\nu$ is of the form
$\eta + \alpha$, where $\eta$ is distributed
according to $\mu$ and $\alpha$ independently according to $\nu$.

Suppose $\eta+\alpha$ can be stabilized, then
we can write
\be\label{een}
\eta_V + \alpha_V - \Delta_V m^1_V = \xi^1_V
\ee
with $m^1_V\uparrow m^1_{\Zd}$ as $V\uparrow\Zd$. We define $m^{2,V}\in\N^{\Zd}$ by
\be\label{twee}
\eta + \alpha^0_V - \Delta m^{2,V} = \xi^{2,V}
\ee
where $\alpha_V^0 : \Z^d\to \N$ is defined $\alpha_V^0 (x) = \alpha (x) \ind_{x\in V}$.
In words this means that we add according to $\alpha$ only in the finite volume $V$ but
we {\em stabilize in infinite volume}. The fact that $m^{2,V}$ is finite
follows from the fact that the addition operators $a_x$ and
finite products of these are well-defined in infinite
volume on $\mu$ almost every configuration. Since for $W \supset V$
\be
\alpha^0_V\leq \alpha^0_W
\ee
and $m^1_V$ does not diverge, it is clear that $m^{2,V}$ is well-defined, by approximating
the equation (\ref{twee}) in growing volumes. Moreover, for $\la\subset \Z^d$ fixed, it is also clear
that $(m^{2,V})_\la$ and $ (m^1_V)_\la$ will coincide for $V\supset V_0$ large enough. Otherwise, the
stabilization of $\eta_V +\alpha_V$ would require additional topplings in $\la$ for infinitely many
$V$'s, which clearly contradicts that $m^1_V$ converges (and hence remains bounded). But then we have
that for $V$ large enough, $(\xi^1_V)_\la$ and $\xi^{2,V}_\la$ coincide. For any $V$, the distribution of
$\xi^{2,V}$ is $\mu$, because $\mu$ is stationary under the infinite volume addition operators. Therefore,
we conclude that the limit $\lim_{V} \xi^1_V = \lim_{V} \xi^2_V$ is distributed according to $\mu$.
Hence, passing to the limit $V\uparrow\Z^d$ in (\ref{twee}) we obtain
\be\label{drie}
\eta + \alpha - \Delta m = \xi
\ee
where $\eta$ and $\xi$ have the same distribution $\mu$, and where $m\in\N^{\Zd}$. Let
$\{X_n,n\in\N\}$ be simple random walk starting at the origin. Then for any function
$f:\Z^d\to\R$,
\be
M_n = f(X_n) - f(X_0) - \frac1{2d}\sum_{i=1}^{n-1} (-\Delta) f(X_i)
\ee
is a mean zero martingale. Applying this identity with $f(x) = m_x$, using (\ref{drie}) gives that
\be
M_n= m(X_n) -m(X_0) + \sum_{i=1}^{n-1}(\eta(X_i) + \alpha (X_i) -\xi (X_i))
\ee
is a mean zero martingale
(w.r.t.\ the filtration $\fe_n =\si(X_r: 0\leq n)$, so $\eta$ and $\xi$ are
{\em fixed} here).

.
This gives, upon taking expectation over the random walk
\be\label{marteq}
\frac{1}{n}\left(\E_0(m(X_n)) - m(0)\right) = \frac1n\E_0\left(\sum_{i=1}^{n-1}(\xi (X_i)-\eta(X_i) -\alpha (X_i))\right)
\ee
Using now that $m(0) <\infty$ by assumption, the ergodicity of $\mu$ and $\nu$, and the fact that both
$\xi$ and $\eta$ have distribution $\mu$, we obtain
from (\ref{marteq}) upon taking the limit
$n\to \infty$ that
\be\label{franken}
0\leq
\lim_{n\to\infty} \frac{1}{n}\left(\E_0(m(X_n)) - m(0)\right)
=
\lim_{n\to\infty}\E_0\left(\frac1n\sum_{i=1}^{n-1}(\xi (X_i)-\eta(X_i) -\alpha (X_i))\right)
= -\alpha
\ee
which is a contradiction.
\epr

In dimension $d=1$ the situation is simpler, see  \cite{quant} for the proof.
\bt
A stationary and ergodic measure $\nu$ on $\N^\Z$ such that
$\nu (\eta(0)) <2$ is stabilizable. If on the contrary
$\nu (\eta(0)) >2$, then $\nu$ is not stabilizable.
\et
\br
For $\nu (\eta (0))= 2$ one can have both stabilizability and
non-stabilizability: e.g.
the configuration $313131313\ldots$ and its shift $13131313\ldots$
are not stabilizable.
\er

\section{Constructive example}

In this section we present an explicit example of an addition which leads to infinitely many topplings at the origin in the
limit $V\uparrow\Zd$. It settles point 2 of the conjecture, even in the case when
the UMRC is not canonical (in particular for $d\leq 4$), and shows that
point 1 is not true in the generality of stationary and ergodic probability measures on height
configurations.

Before presenting the example, let us recall the following fact about the abelian sandpile model, see e.g.\ \cite{IP}.
Consider the abelian sandpile model in a finite volume $V\subset\Z^d$. Start from
a recurrent stable height configuration $\eta: V\to \{ 1,\ldots, 2d\}$. Add on each boundary
site $x\in\partial V$ as many grains as there are ``lacking neighbors", i.e.,
the number of grains added at $x$ equals $\lambda_V (x) =|\{ y\in\Z^d: |y-x|=1, y\not\in V\}|$. Then, upon stabilization, each
site will topple exactly once and the recurrent configuration $\eta$ will remain unaltered. If $V$
is a rectangle, and a stable height configuration in $V$ is recurrent, then this ``special addition"
consists of adding two grains to the corner sites and one grain to the other boundary sites. If
$\eta: V\to \{ 1,\ldots,2d\}$ is recurrent in $V$, and $W\subset V$, then the restriction
$\eta_W$ is recurrent in $W$. Therefore, by abelianness, if we add to each boundary site $x$ of $W$
{\em at least}
a number grains
$\alpha^W_x= \{y\in Z^d: |y-x|=1, y\not\in W\}$, then, upon stabilization, each site in $W$ will topple
{\em at least} once.

We can now present our example in the case $d=2$; the generalization to $d\not= 2$ is obvious.
Let $\omega, \omega'$ be independent and distributed according to a Bernoulli measure $\pee_p$ on $\{0,1\}^{\Z}$, with
$\pee_p (\omega(x)=1)=p$.
Consider the following two dimensional random field
$\zeta (x,y) = \omega (x) + \omega' (y)$. This is what we are going to add to
a recurrent configuration. In words, if for $x\in\Z, \omega (x) =1$, then we add one grain to
each lattice site of the vertical line $\{ (x,y), y\in\Z\}$, and if $\omega' (y) =1$ then
we add one grain to each lattice site of the horizontal line $\{ (x,y) :x\in\Z\}$.
If we add according to $\zeta$, then there are almost surely
infinitely many rectangle $R_1,\ldots R_n,\ldots$ surrounding the
origin with
corner sites where we add two grains and other boundary sites where we add at least one grain.

If we add such a configuration $\zeta$ to any recurrent configuration $\eta$, then we have
that the number of topplings at the origin in the finite volume $V$ is at least the number of rectangles $R_i$ that are inside $V$. Indeed, upon addition
according to $\zeta$ on such a rectangle, every site inside the rectangle will topple
at least once.

Therefore the distribution $\mu_p$ of $\eta+ \zeta$ where $\eta$ is drawn from
the UMRC $\mu$, is not stabilizable. Since we can choose $p$ arbitrary close to zero, any density
$\rho\in(\rho_c,\rho_c +2)$ can be attained by $\mu_p$.

To show that we can actually get below $\rho_c$, remember that the fact that the number of
topplings inside $V$ is at least the number of rectangles $R_i$ does only depend on the fact that
the configuration to which we add is recurrent, and not on the fact that the configuration
is chosen from a particular distribution.

Therefore,
consider a translation invariant probability measure  $\mu'$ concentrating
on {\em a subset} $\re'$ of $\re$.
Then with the same reasoning, the distribution $\mu'_p$ of $\eta+ \zeta$ where $\eta$ is drawn from
the  $\mu'$, is not stabilizable. Consider therefore $\mu'$ to be a weak limit point of the uniform
measures on {\em minimal recurrent configurations}, where ``minima'' is in the sense of
the pointwise ordering of configurations. Then
the distribution $\mu'_p$ has its expectation $\mu'_p (\eta (0))$ arbitrary close to
$\mu'(\eta(0)) $. This number is strictly less than the critical density $\rho_c$. Indeed,
it is at most $3$, because the all three configuration is recurrent, and
$\rho_c >3$, see e.g., \cite{Priezzhev}.

 This shows that point 1 of the conjecture cannot hold in that generality.

Combining our results so far with proposition 2.4 from \cite{quant}, we conclude
\bt
Let $\rho^+_c= \rho^-_c$ be as in lemma \ref{monlem}. Then
\be
\rho^+_c= \inf \{ \nu (\eta (0)): \nu\  \mbox{ is translation invariant and}\ \nu (\re) =1\}.
\ee
For any dimension $d$,
$\rho^+_c\leq\rho_c$, and for $d=2$, $\rho^+_c\leq 3 < \rho_c$
\et
\br
We believe that the strict inequality $\rho^+_c <\rho_c$ holds in {\em any} dimension $d>1$, but this is
an open problem as far as we know. In $d=1$ we have $\rho^+_c= 2 = \rho_c$, but this is an exceptional
case where almost all recurrent configurations are minimal recurrent.
\er

\section{Other notions of stabilizability}
\subsection{Stabilization in infinite volume}
\bd
A configuration $\eta\in\N^{\Zd}$ is called weakly stabilizable if there
exists $m\in\N^{\Zd}$ and $\xi\in\Omega = \{ 1,\ldots,2d\}^{\Zd}$ such that
\be
\eta-\Delta m = \xi
\ee
\ed
It is clear that if $\eta$ is stabilizable, then it is weakly stabilizable and
we can choose $m=\lim_{V\uparrow\Zd} m_V$. However it is not clear whether there exist
unstable configurations which can be stabilized {\em directly in infinite volume}
but which satisfy $\lim_{V\uparrow\Zd} m_V (0) =\infty$, i.e., the infinite volume
toppling numbers are not obtained as the limit of toppling numbers in larger and larger
volumes. In the following proposition we prove that a measure $\nu$ with
$\nu (\eta (x))> 2d$ for all $x\in\Zd$ cannot be weakly stabilized. This means in words that
mass cannot be ``swept away" to infinity.

The following example shows that the opposite, importing mass from infinity, is not impossible. Consider  $f:\Z^2\to\N$:
\[
f(x,y)= x^2+ y^2
\]
then $\Delta f = -4$ and hence, for example,
\be\label{exa}
\overline{6} = \overline{2} -\Delta f
\ee
where $\overline{6}$ (resp.\ $\overline{2}$) denotes the configuration with height $6$ (resp.\ 2) at every site.
However,
\[
\lim_{V\uparrow\Zd} \Delta_{V}^{-1} (\overline{6}-\overline{2}) = \infty
\]
so this ``infinite volume toppling" cannot be obtained as a limit of finite volume
topplings. Notice however that ``toppling" according to $f$ is not ``legal" in the following sense:
we cannot find an order of topplings such that, performed in this order, only unstable sites topple
and at the end every site has toppled $x^2 +y^2$ times. The point of (\ref{exa}) is that the
equality
\be\label{blub}
\eta = \xi -\Delta m
\ee
for some $m\in\N^{\Zd}$
does not imply that the densities of $\eta$ and $\xi$ are equal. However as we will see later, the equality
in (\ref{blub}) {\em does imply } that the density of $\eta$ is {\em larger or equal} than that of $\xi$.

In the following proposition we show point 3 of the conjecture for weak stabilizability.
\bp
Let $\nu$ be stationary ergodic such that $\nu (\eta (0)) > 2d$. Then
$\nu$ is not weakly stabilizable.
\ep
\bpr
Suppose that there exist $m\in\N^{\Zd}$ such that
\be\label{altijd}
\eta -\Delta m = \xi
\ee
with $\xi$ stable and $\eta$ a sample from $\nu$.
Let $X_n$ be the position of simple random walk starting at the origin at time $n$.
>From (\ref{altijd}) it follows that
\be
m(X_n)- m(0) - \frac {1}{2d} \sum_{k=1}^{n-1}(\xi (X_k) -\eta (X_k))
\ee
is a mean-zero martingale.
Therefore taking expectations w.r.t. the random walk
\be
\frac{1}{n}\left(\E_0 (m(X_n)) - m(0)) \right)= \frac{1}{n}\E_0 \left(\frac {1}{2d}\sum_{k=1}^{n-1}(\xi (X_k) -\eta (X_k))\right)
\ee
By stability of $\xi$
\be
 \frac1n \left(\sum_{k=1}^{n-1}(\xi (X_k) -\eta (X_k))\right) \leq 2d - \frac1n\sum_{k=1}^{n-1}(\eta (X_k))
\ee
Therefore, using dominated convergence and $\nu (\eta(0)) = 2d + \delta$,
\beq
0& \leq & \liminf_{n\to\infty}\frac{1}{n}\left(\E_0 (m(X_n)) - m(0)\right)\nonumber\\
& \leq & \limsup_{n\to\infty}\frac1n \E_0\left(\frac {1}{2d}\sum_{k=1}^{n-1}(\xi (X_k) -\eta (X_k))\right)
\nonumber\\
&\leq &
1-\E_0\left(\frac {1}{2d}\lim_{n\to\infty} \frac1n\sum_{k=1}^{n-1}\eta (X_k)\right) <-\delta
\eeq
which is a contradiction.

\epr

\subsection{Activated random walkers system and stabilizability at low density}
Dickman proposes in \cite{dick} the following mechanism of stabilization. Consider
a configuration $\eta\in\haa$. To each $x\in\Zd$ is associated a
Poisson process $N^x_t$, for $x\not= y$ these processes are independent.
On the event times of $N^x_t$ a site topples if it is
unstable (the random walkers are ``activated"), otherwise nothing happens. This means that after time $t$ the
configuration $\eta$ evolves towards $\eta_t$ according to a Markov process
with generator
\[
L f(\eta ) = \sum_{x\in\Zd} \ind_{\eta (x) > 2d}\left( f(\eta - \Delta_{x,\cdot})- f(\eta)\right)
\]
One says now that the configuration is stabilizable by this process if
for any $x\in\Zd$ the value $\eta_t (x)$ jumps only a finite number of times. One can write
the configuration $\eta_t$ as
\be\label{dikstab}
\eta_t = \eta_0- \Delta n^t_\eta
\ee
where $n^t_\eta$ is the vector collecting at each site the number of topplings
at $x$ in $[0,t]$. If $\eta$ is distributed according to a translation invariant
probability measure $\nu$ on $\N^{\Zd}$ then so is $n^t_\eta$ under the joint
measure $\nu\times \pee$ where $\pee$ is the distribution of the
Poisson processes $N^t_x, x\in\Zd$. Moreover $n^t_\eta(x)\leq N^x_t$ by definition and hence
\[
\E(\eta_t(x)) = \E (\eta_0 (x))
\]
i.e., this process conserves the density.

\bl\label{stablem}
A configuration is stabilizable by the process with generator $L$ if and
only if it is stabilizable (in the sense of definition \ref{stabdef}).
\el
\bpr
Suppose $\eta$ is stabilizable by the process with generator $L$.
Consider then the generator
\[
L_V f(\eta ) = \sum_{x\in V} \ind_{\eta (x) > 2d}\left( f(\eta - \Delta^V_{x,\cdot})- f(\eta)\right)
\]
corresponding to toppling inside $V$ only, according to the finite volume toppling
matrix, on the event times of the Poisson process $N^t_x$.
Call $n^t_V(x,\eta) $ the number of updates of $x$ in $[0,t]$. It is
easy to see that $n^t_V (x,\eta)\uparrow n^t_\eta (x)$ as $V\uparrow\Zd$, where
$n^t_\eta$ is defined above. By assumption,
for any $x$ there exists $t^V_x(\eta)$ such that for any $t\geq t^V_x(\eta)$
$n^t_V (x,\eta)=n^{t^V_x(\eta)}_V (x,\eta)$, and moreover $t^V_x(\eta)\uparrow t_x(\eta)$ for which
$n^t_\eta (x)= n^{t_x{\eta}}_\eta (x)$ for any $t>t_x(\eta)$. Therefore in definition
\ref{stabdef} we can identify
\[
m_V (x)=n^{t^V_x(\eta)}_V (x,\eta)
\]
and
\[
m (x)=n^{t_x{\eta}}_\eta (x)
\]
Suppose that $\eta$ is stabilizable in the sense of definition \ref{stabdef}.
Then, clearly, in finite volume we have the equality
\[
m_V (x)=n^{t^V_x(\eta)}_V (x,\eta)
\]
Since $m_V\uparrow m$ we have
\[
\sup_{V\subset\Zd}n^{t^V_x(\eta)}_V (x,\eta) <\infty
\]
and hence
\[
t_x(\eta) = \sup_{V\subset\Zd} t^V_x (\eta)
\]
is finite $\pee$ almost surely. Now pick $t>t_x (\eta)$. Then
\be\label{fifi}
n_t (x,\eta) =\lim_{V\uparrow\Zd} n^t_V (x,\eta)
=\lim_{V\uparrow\Zd}
n^{t^V_x(\eta)}_V (x,\eta)
\ee
where in the second step we used that the processes with generator $L_V$ converge
to the process with generator $L$ weakly on path space. Indeed for
any local function
\[
\lim_{V\uparrow \Zd} L_V (f) = L(f)
\]
So the convergence of the processes follows from the Trotter-Kurtz theorem.
The right hand side of
(\ref{fifi})
does not depend on $t$ anymore. Hence $\eta$ is stabilizable
by the process with generator $L$.
\epr

In \cite{quant} the authors prove that there exists $\rho'_c>0$ such that
if $\nu (\eta (0))< \rho'_c$, then $\nu$ is stabilizable by the process with generator
$L$. $\rho'_c$ is the density of ``minimal recurrent configurations".
The following theorem is then an immediate consequence.
\bt
There exists $\rho'_c>0$ such that if $\nu$ is any stationary ergodic
probability measure on $\haa$ with $\nu (\eta (0))<\rho'_c$, then $\nu$ is stabilizable.
\et
\bpr
Combine lemma \ref{stablem} with proposition 2.4 from \cite{quant}.
\epr
\section{Metastable measures}
Suppose that $\nu\geq \mu$, and $\nu (\eta (0)) > \mu (\eta (0))$, i.e., $\nu$ has
a strictly higher density than $\mu$, and stochastically dominates $\mu$. In that situation
$\nu$ can still concentrate on stable configurations, and hence be stabilizable.
One feels however that such a measure is ``on the brink of non-stabilizability".
This is formalized in the following definition.
\bd
A measure $\nu$ is called metastable if it is stabilizable and if
$\nu\oplus \delta_0$ is not stabilizable with non zero probability, i.e.,
$\nu\oplus\delta_0 (\s) <1$.
\ed
In words this means that upon stabilizing $\nu\oplus\delta_0$ in volume $V$, with positive probability
the number of topplings $m_V (0)$ diverges as $V\uparrow\Zd$.
The simplest example of a metastable measure is the measure concentrating on the maximal
stable configuration $\nu= \delta_{2d}$.
The following theorem shows that there are other non-trivial metastable measures.
\bt\label{metstabthm}
Suppose that $\nu$ is a stationary and ergodic probability measure
on $\Omega$, concentrating on the set of recurrent configurations $\re$.
Define $I_\eta (x) = \ind_{\eta (x) =2d}$ and call $\tilde{\nu}$ the distribution
of $I_\eta$. Suppose that $\tilde\nu$ dominates a bernoulli measure
$\pee_p$ with $p$ sufficiently close to one such that
the $1$'s percolate and the zeros do not percolate. Then $\nu$ is metastable.
\et
Before giving the proof, we need the concept of ``wave", see e.g.\ \cite{IP}.
Consider a stable height configuration in a finite volume $V$. Add to $x$ and
topple $x$ once (if necessary) and perform all the necessary topplings except
a new toppling at $x$. This is called the first wave. The support of the wave
is the set of sites that toppled (it is easy to see that during a wave all sites
topple at most once). If $x$ is still unstable,
then iterate the same procedure; this gives the second wave etc. For the proof of
theorem \ref{metstabthm}, we need that the support of a wave is simply connected, i.e.,
contains no holes. For the sake of completeness, we give a precise definition of
this and prove this property of the wave, already formulated in \cite{IP}.
\bd
A subset $V\subset\Z^d$ is called simply connected
if the set obtained by ``filling the squares", i.e.,
the set
\[
\cup_{i\in V} (i+ [-1/2,1/2]^d)
\]
is simply connected.
\ed
\bl
Let $\eta\in\re_V$ be a recurrent configuration. Then the support of any wave
is simply connected.
\el
\bpr
Suppose the support of the
first wave contains a maximal (in the sense of inclusion) hole $\emptyset\not= H\subset V$. By hypothesis, all the neighbors of $H$ in
$V$ which do not belong to $H$ have toppled once. The toppling of the outer boundary gives
an addition to the inner boundary equal to $\lambda_H(x)$ at site $x$, where
$\lambda_H (x)$ is the number of neighbors of $x$ in $V$, not belonging to $H$. But addition
of this to $\eta_H$ leads to one toppling at each site $x\in H$, by recurrence of $\eta_H$.
This is a contradiction because we supposed that $H\not=\emptyset$, and $H$ is not contained in the
support of the wave. After the first wave the restriction of the configuration
to the volume $V\setminus x$ is recurrent
{\em i.e., recurrent in that volume $V\setminus x$}. Suppose that the second
wave contains a hole, then this hole has to be a subset of $V\setminus x$
(because $x$ is contained in the wave by definition), and arguing as
before, one sees that the subconfiguration $\eta_H$ cannot be recurrent.
\epr
We can now give the proof of theorem (\ref{metstabthm}).
\bpr
The idea of the proof is the following. Suppose we have a ``sea" of height $2d$ and
``islands" of other heights, and such that the configuration is recurrent.
Suppose the origin belongs to the sea, and we add a grain at the origin. The first wave must be a simply connected
subset of $\Zd$ because the configuration is recurrent. It is clear that the ``sea" of height $2d$ is part of the wave, and therefore
every site is contained in the wave (because if an island is not contained then the wave
would not be simply connected). So in the first wave every site topples exactly once, but this implies
that the resulting configuration is exactly the same. Hence we have infinitely many waves.

Let us now formalize this. For a given configuration $\eta$ (distributed according
to $\nu$) a volume $V\subset \Zd$ is called ``a lake with islands" if
all the boundary sites of $V$ have height $2d$, and if from the origin
there is a path along sites having height $2d$ to the boundary. From the fact that the zeros do not percolate,
and $1$'s do percolate
it follows that with positive probability, the origin has height $2d$ and is in infinitely many nested lakes, i.e.
$V_1 \subset V_2,\ldots ,V_n\ldots$, with for $i \neq j,\ \partial V_i \cap \partial V_j = \emptyset$. Consider a configuration from that event, consider a volume
$V\supseteq V_n$, add a grain at the origin and stabilize the configuration in $V$. In the first wave all sites will topple
once because islands not contained in the wave would contain forbidden subconfigurations, which
is impossible since the configuration is recurrent. After the first wave, the only sites that
change height are on the boundary of $V_n$. Therefore, the origin is still unstable and a second
wave must start. The sites included in this wave will contain the set of sites included in the first
wave needed to stabilize $\eta_{V_{n-1}}$, but this set, with the same argument, is at least
$V_{n-1}$. Continuing like this, one sees that at least $n$ waves are needed for the
stabilization of $\eta$ inside $V$. Since with positive probability we find
infinitely many lakes containing the origin, the number of topplings is diverging in the
limit $V\uparrow\Zd$ with positive probability, which is what we wanted to prove.
\epr

\br
We believe but cannot prove that {\em any} translation invariant probability measure $\nu\in\caP (\haa)$ with
$\nu (\eta (0)) \in (\rho_c, 2d)$ is either metastable or not stabilizable.
\er
\section{Appendix}
In this appendix we prove the ergodicity of the scenery process, used at several places in this,
e.g.\ in (\ref{franken}).
\bp
Let $\Omega = \N^{\Zd}$ (or any space on which translations act).
Suppose that $\mu$ is a stationary and ergodic probability measure, and
$X_n,n\in\N$ is symmetric nearest neighbor random walk on $\Zd$. Then, if initially
$\eta$ is distributed according to $\mu$,
the process
$\tau_{X_n} \eta$ defined by
\[
\tau_{X_n} \eta (x) = \eta (x+ X_n)
\]
is a stationary ergodic (in time $n$) Markov process.
\ep
\bpr
Let $S_d$ denote the set of unit vectors of $\Zd$.
The process $\tau_{X_n}\eta$ is clearly a Markov process with transition
operator
\[
Pf (\eta) = \frac{1}{2d}\sum_{e\in S} f(\tau_e\eta)
\]
To prove ergodicity (in time $n$) of the Markov process, we have to show that if $f$ is a bounded
measurable function with $Pf=f$, then
$f$ is constant $\mu$-almost-surely (see e.g. \cite{Rosen}). So suppose that $Pf=f$, then, by translation
invariance of $\mu$,
\[
\int f(Pf-f)d\mu = -\frac{1}{4d} \sum_{e\in S_d} \int(\tau_e f- f)^2 d\mu =0
\]
Hence $\tau_e f= f$, $\mu$ almost surely for all unit vectors, and hence
$\tau_x f = f$ $\mu$ almost surely for all $x\in\Zd$. By ergodicity of $\mu$
(under translations), this implies $f=\int f\mu$ $\mu$-almost surely.
\epr
Finally, we prove a fact about recurrent configurations that we used in our constructive example.

Define two (possibly unstable) height configurations $\eta, \xi$ on $V$ equivalent
if there exists $m_V: V\to \Z$ such that
$\eta- \xi = \Delta_V m_V$. It is well-known that
every equivalence class of this relation contains exactly one recurrent configuration. So if
$\eta$ is a (possibly unstable) height configuration which upon stabilization
yields  a recurrent configuration $\xi$, then the relation
\be\label{topot}
\eta - \Delta_V m_V = \xi
\ee
simply means that stabilization of $\eta$ requires $m_V (x)$ topplings at site
$x\in V$ and yields $\xi$ as end result.
\bp
Suppose that $V\subset\Z^2$ is a rectangle and $\eta$ is a recurrent
configuration in $V$. Then upon addition of $2$ grains to the
corner sites and $1$ grain to all other boundary sites, every site will
topple once and the resulting configuration remains unaltered.
\ep
\bpr
Let $\overline{1}$ denote the column indexed by $x\in V$ of all ones, then
$(\Delta_V \overline{1})_x $ equals $4$ minus the number of neighbors of $x$
in $V$. Therefore, the simple identity
\[
\eta + \Delta_V\overline{1} - \Delta_V\overline{1} = \eta
\]
shows that if $\eta$ is a recurrent configuration, addition of $(\Delta_V \overline{1})_x $
reproduces $\eta$, and makes each site topple once (cf.\ (\ref{topot})).

\epr

{\bf Acknowledgments.} We thank Corrie Quant and Ronald Meester for interesting discussions and
suggestions.

\end{document}